\documentclass{ws-p9-75x6-50}

\begin{document}

\title{Broken Symmetry and Josephson-like Tunneling in Quantum Hall Bilayers}

\author{S. M. Girvin}

\address{Department of Physics\\
Indiana University\\
Bloomington, IN 47405 USA}




\maketitle

\abstracts{I review recent novel experimental and theoretical advances 
in the physics of quantum Hall effect bilayers. Of particular
interest is a broken symmetry state which optimizes correlations 
by putting the electrons into a coherent superposition of the 
two different layers.}

The various quantum Hall effects are among the most remarkable
many-body phenomena discovered in the second half of the
twentieth century.
\cite{prangegirvin,tapash,perspectivesbook,leshouches} The
fractional effect has yielded fractional charge, spin and
statistics, as well as unprecedented order parameters.
\cite{ODLRO}  There are beautiful connections with a variety of
different topological and conformal field theories of interest in
nuclear and high energy physics.

The quantum Hall effect (QHE) takes place in a two-dimensional
electron gas formed in a quantum well in a semiconductor host
material and subjected to a very high magnetic field.  In essence
it is a result of a commensuration between the number of
electrons, $N$, and the number of flux quanta, $N_\Phi$, in the
applied magnetic field.  The electrons condense into distinct and
highly non-trivial ground states (`vacua') formed at each
rational fractional value of the filling factor $\nu \equiv
N/N_\Phi$.

The essential feature of (most) of these exotic states is the
existence of an excitation gap.  The electron fluid is {\em
incompressible} and flows rigidly past obstacles (impurities in
the sample) with no dissipation. 
A weak external electric field
will cause the fluid to move, but the excitation gap prevents the
fluid from absorbing any energy from the electric field. Hence
the current flow must be exactly at right angles to the field 
and the conductivity tensor takes the remarkable universal form
\be\sigma^{xx} = \sigma^{yy} = 0;\,\,\,\,\,
\sigma^{xy} = -\sigma^{yx} = \nu\frac{e^2}{h}.
\ee
Ironically, this ideal behavior occurs because of imperfections
and disorder in the samples which localize topological defects
(vortices) whose motion would otherwise dissipate energy.  In a
two-dimensional superconductor, such vortices undergo a
confinement phase transition at the Kosterlitz-Thouless
temperature and dissipation ceases.  In most cases in the QHE, an
analog of the Anderson-Higgs mechanism  causes the vortices to be
deconfined \cite{ODLRO} so that dissipation is strictly zero only
at zero temperature.  In practice, values of
$\sigma^{xx}/\sigma^{xy}$ as small as $10^{-13}$ are not
difficult to obtain at dilution refrigerator tempertures.

Recent technological progress in molecular beam epitaxy
techniques has led to the ability to produce pairs of closely spaced
two-dimensional electron gases. 
Strong correlations between the electrons in different
layers lead a great deal of completely
new physics involving spontaneous interlayer phase
coherence. \cite{fertig,wenzee,perspectivesbook,kyangprl,kmoonprb,%
kyangprb,Schlieman,ady1PRL,ady2PRL,leonleoPRL,fogler,chetan,yogesh,%
veillette,burkov}

As we will discuss below, this is the first example of a QHE
system with a finite-temperature phase transition.  This
transition is in fact a Kosterlitz-Thouless transition into a
broken symmetry state which is closely analogous to that of a 2D
superfluid.  Recent remarkable tunneling experiments
\cite{jpetunnel1,jpegoldstone} have observed something closely
akin to the Josephson effect in superconducting tunnel junctions
and have measured the dispersion of the superfluid Goldstone mode.

We begin with the simplest example of the
integer QHE in a single layer system of spinless electrons 
at $\nu=1$. The strong magnetic field
quantizes the kinetic energy into discrete Landau levels
\cite{leshouches} separated in energy by the cyclotron energy
$\hbar \omega_{\rm c} \sim 100$K.    Each level has  a {\em
macroscopic} degeneracy equal to $N_\Phi$.  This degeneracy in the
kinetic energy means that interactions are enormously important
and have non-perturbative effects at fractional filling factors.
However for $\nu=1$, every state of the lowest Landau level (LLL)
is occupied and, since there is a large kinetic energy gap to the
next Landau level, interactions are (relatively) unimportant.  It
is this gap which makes the system incompressible. Since the
lowest LLL is completely full, 
the state is a simple Slater determinant.  In the Landau
gauge \cite{leshouches} this can be written in second quantized
form 
$
 |\Psi\rangle = \prod_k c^\dagger_k |0\rangle 
$
 where
$k$ labels the set of single-particle states.  In first-quantized
form the state is most easily expressed in the symmetric gauge
\cite{leshouches} $ \Psi(z_1,z_2,\ldots,z_N) = \prod_{i<j}^N
(z_i - z_j) e^{- \frac{1}{4}\sum_m |z_m|^2} $ where $z_j \equiv
(x_j + i y_j)/\ell$ is a dimensionless complex number
representing the 2D position vector of the $j$th particle in
units of the magnetic length $\ell$.  The vandermonde polynomial
factor in this Laughlin state is totally antisymmetric and is
equivalent to a single Slater determinant filling all the
orbitals in the LLL.

So far we have been ignoring the spin of the electrons.  Various
solid state effects make the Zeeman splitting much smaller than
the LL splitting and so the degeneracy of each LL is effectively
doubled when we include spin.  Thus interactions turn out to be
much more important \cite{sondhiskyrme} than we have been naively
assuming. At $\nu=1$ the Coulomb interaction makes the spins
spontaneously align into a very simple maximally ferromagnetic
state \cite{leshouches} \be \Psi(z_1,z_2,\ldots,z_N) =
\prod_{i<j}^N (z_i - z_j) e^{- \frac{1}{4}\sum_m |z_m|^2}
|\uparrow\uparrow\uparrow\uparrow\ldots\uparrow\rangle . \ee In
this state the spin part of the wave function is fully symmetric
and so the spatial part must be the same fully antisymmetric wave
function considered above. This wave function vanishes whenever
any two electrons approach each other and thus optimizes the
Coulomb exchange energy. Because (unlike an ordinary
ferromagnetic metal) there is no kinetic energy cost to aligning
the spins, the polarization is 100\%. In the subsequent
discussion we will assume that the weak Zeeman splitting combined
with strong Coulomb exchange has frozen out the spin degree of
freedom.  This simplifying assumption is not necessarily valid in
real systems at low magnetic fields however.

We turn now to the case of a QHE bilayer at {\em total} filling
factor $\nu=1$, that is, filling $1/2$ in each layer.  In nuclear
physics the strong interaction between nucleons is largely
independent of whether they are neutrons or protons.  In that
case it proves useful to define an isospin variable in which the
up and down states of this new spin-1/2 degree of freedom
represent the proton and neutron.  Similarly here, if the layer
spacing $d$ is small compared to the electron spacing (which is a
few times the magnetic length $\ell$), then the Coulomb
interaction between the particles is nearly independent of which
layer they are in. If we define an isospin or pseudospin which
labels the layer index, then the interactions are nearly
rotationally invariant in the pseudospin space.   For $d=0$ there
is an exact SU(2) invariance and the ground state at $\nu=1$ is
identical to the ferromagnetic case discussed above for real spins
\be |\Psi\rangle = \prod_k c^\dagger_{k\uparrow} |0\rangle. \ee
The only difference is that the up arrow now represents an
electron being in the upper layer.  The system is a pseudospin
ferromagnet for precisely the same reason that it is a real spin
ferromagnet--this state optimizes the Coulomb exchange energy by
making the spatial part of the wave function fully
antisymmetric.  This state has total pseudospin $S =
\frac{N}{2}$.  The particular realization above has
$S^z=\frac{N}{2}$, but there are a total of $2S+1$ degenerate
states with all possible pseudospin orientations.

Consider now what happens for small but finite layer separation
$d$. In this limit the Coulomb interaction between electrons in
the same layer is slightly stronger than for electrons in
different layers.  Thus the interactions are no longer pseudospin
rotationally invariant.  For small $d$ the main effect of this is
not to change the wave functions of the eigenstates described
above but simply to lift their degeneracy. This is accurately
represented by an `easy plane' anisotropy term $H_a$ in the
energy of the pseudospin ferromagnet \be H_a =
\frac{e^2}{2C}(S^z)^2. \ee Because $S^z = (N_\uparrow -
N_\downarrow)/2$ represents the charge imbalance between the two
layers, this term is simply the charging energy of the capacitor
$C$ whose plates are the two electron gases. This weak anisotropy
prefers for the pseudospin magnetization to lie in the xy plane
so that $\langle S^z\rangle = 0$.   A single spin lying in the xy
plane at an angle $\varphi$ with respect to the $x$ axis is a
linear combination of the two up and down basis states: \be
|\rightarrow\rangle = \frac{1}{\sqrt{2}}\left(|\uparrow\rangle +
e^{i\varphi}|\downarrow\rangle\right). \ee Hence the fully
ferromagnetic many-body state with the same orientation is given
by \cite{fertig} \be |\Psi\rangle = \prod_k
\frac{1}{\sqrt{2}}\left(c^\dagger_{k\uparrow} +
e^{i\varphi}c^\dagger_{k\downarrow}\right)|0\rangle
\label{eq:fertig} \ee or equivalently in first quantization \be
\Psi(z_1,z_2,\ldots,z_N) = \prod_{i<j}^N (z_i - z_j) e^{-
\frac{1}{4}\sum_m |z_m|^2}
|\rightarrow\rightarrow\rightarrow\ldots\rightarrow\rangle.
\label{eq:fertig1stquant} \ee This is a very strange state.  Even
though there may be no possibility of tunneling between the two
layers, quantum mechanics allows the existence of states in which
we are uncertain which layer each electron is in.  This state has
this property--it exhibits spontaneous interlayer phase
coherence.  Each electron is in a coherent superposition of the
upper and lower layers, characterized by the phase angle
$\varphi$.

This state represents a broken gauge symmetry much like that in a
superconductor.  A superconductor spontaneously breaks the gauge
symmetry associated with total charge.  The bilayer QHE system is
incompressible and has definite total charge.  However it has
fluctuations in the charge difference between the two layers due
to the uncertainty over which layer each electron is in.  Hence
it breaks the gauge symmetry associated with conservation of the
charge {\em difference} between the two layers. \cite{wenzee} To
understand this, consider the gauge transformation induced by the
unitary operator $ U_-(\theta) = e^{i\frac{\theta}{2}
(N_\uparrow - N_\downarrow)}$.
The effect of this
transformation on the field operators is \be U_-^\dagger\,
c^\dagger_{k\uparrow}\, U_- =
e^{-i\frac{\theta}{2}}c^\dagger_{k\uparrow};\,\,\,\, 
U_-^\dagger\, c^\dagger_{k\downarrow}\, U_- =
e^{+i\frac{\theta}{2}}c^\dagger_{k\downarrow}. \ee The
Hamiltonian is invariant under this U(1) transformation \be
U_-^\dagger(\theta)\, H\, U_-(\theta) = H \ee since \be
[H,(N_\uparrow - N_\downarrow)] =0, \ee in the absence of
tunneling between the layers.  Examination of
Eq.~(\ref{eq:fertig}) however shows that the phase coherent state
is characterized by the non-trivial  order parameter \be
\psi(\vec r) \equiv \langle \Psi^\dagger_\uparrow(\vec r)
\Psi_\downarrow(\vec r)\rangle = \frac{n_0}{2} e^{i\varphi}
\label{eq:orderparameter} \ee where $\Psi^\dagger_\sigma(\vec r)$
creates a particle in layer $\sigma$ at position $\vec r$ and
$n_0 = 1/(2\pi\ell^2)$ is the total density. This order parameter
is {\em not} gauge invariant \be \psi(\vec r) \longrightarrow
\langle U_-^\dagger(\theta)\,\Psi^\dagger_\uparrow(\vec r)
\Psi_\downarrow(\vec r) U_-(\theta)\rangle = e^{i\theta}\psi(\vec
r). \ee Thus the state has less symmetry than the Hamiltonian,
and it spontaneously breaks the U(1) symmetry associated with
conservation of $N_\uparrow - N_\downarrow$.

In a superconductor, the pair field order parameter $ \chi(\vec
r) \equiv \langle \Psi^\dagger_\uparrow(\vec r)
\Psi^\dagger_\downarrow(\vec r) \rangle $ transforms
non-trivially under the gauge transformation associated with
conservation of {\em total} charge $ U_+(\theta) =
e^{i\frac{\theta}{2} (N_\uparrow + N_\downarrow)}$.  The
bilayer order parameter is however invariant under this
transformation.  The order parameter is {\em charge neutral}--it
corresponds to pairing of particles and holes rather than
particles and particles.  Because of the charge neutrality, the
pairs can condense despite the presence of the strong magnetic
field.  In contrast, the order parameter of a superconductor
would be filled with vortices by the magnetic field which thus
discourages condensation.

Note that we do not have the situation of a particle in one
particular layer bound to a hole in the other layer.  Rather we
have one particle in each spatial orbital but we are uncertain
which layer it is in.  That is, the particle is in one layer and
the hole in the other, but we do not know which is which. We see
from the first quantized form of the state in
Eq.~(\ref{eq:fertig1stquant}) that there is a zero of the wave
function whenever any two particles approach each other. Whether
they are in the same or different layers does not matter.  Thus
if the particle is in the upper layer, there is guaranteed to be a
correlation hole directly underneath it in the other layer and
vice versa.  This explains the good correlation energy for this
state.

\begin{figure}[ht]
 \centerline{ 
\epsfysize=3.00in \epsffile{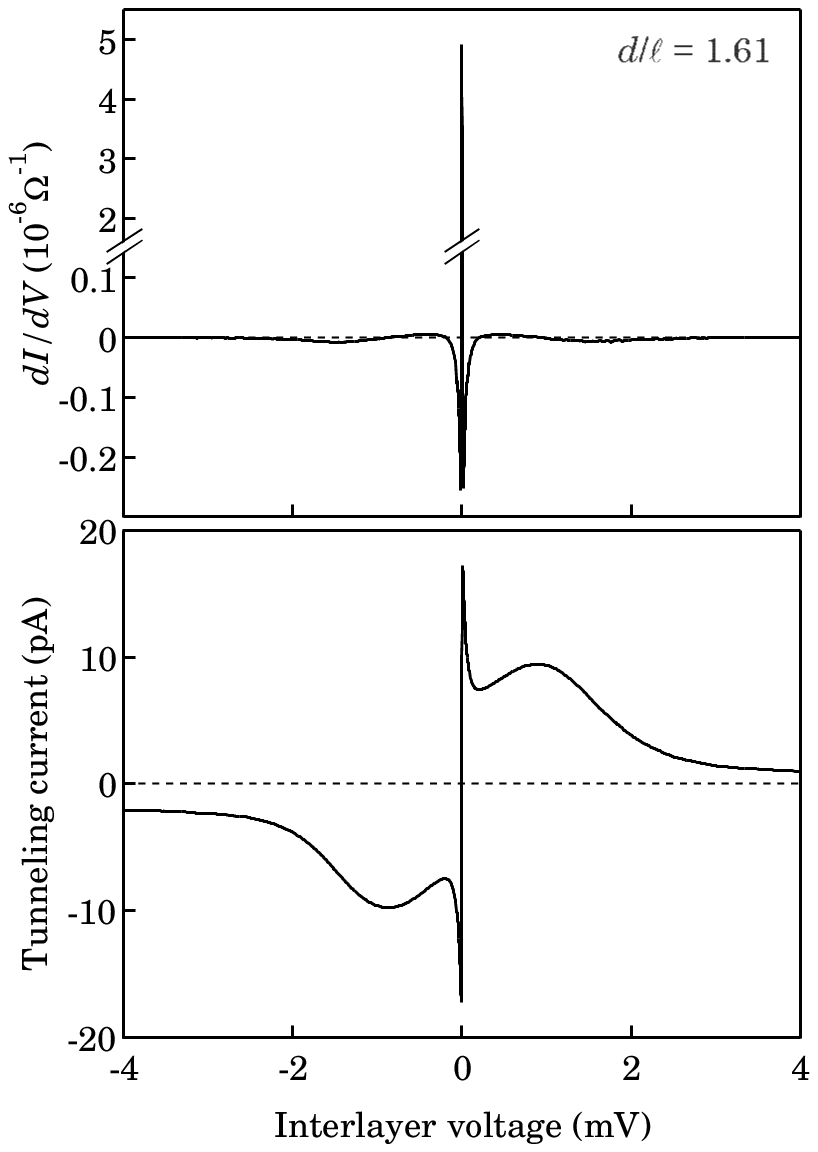}    
 \epsfysize=2.95in \epsffile{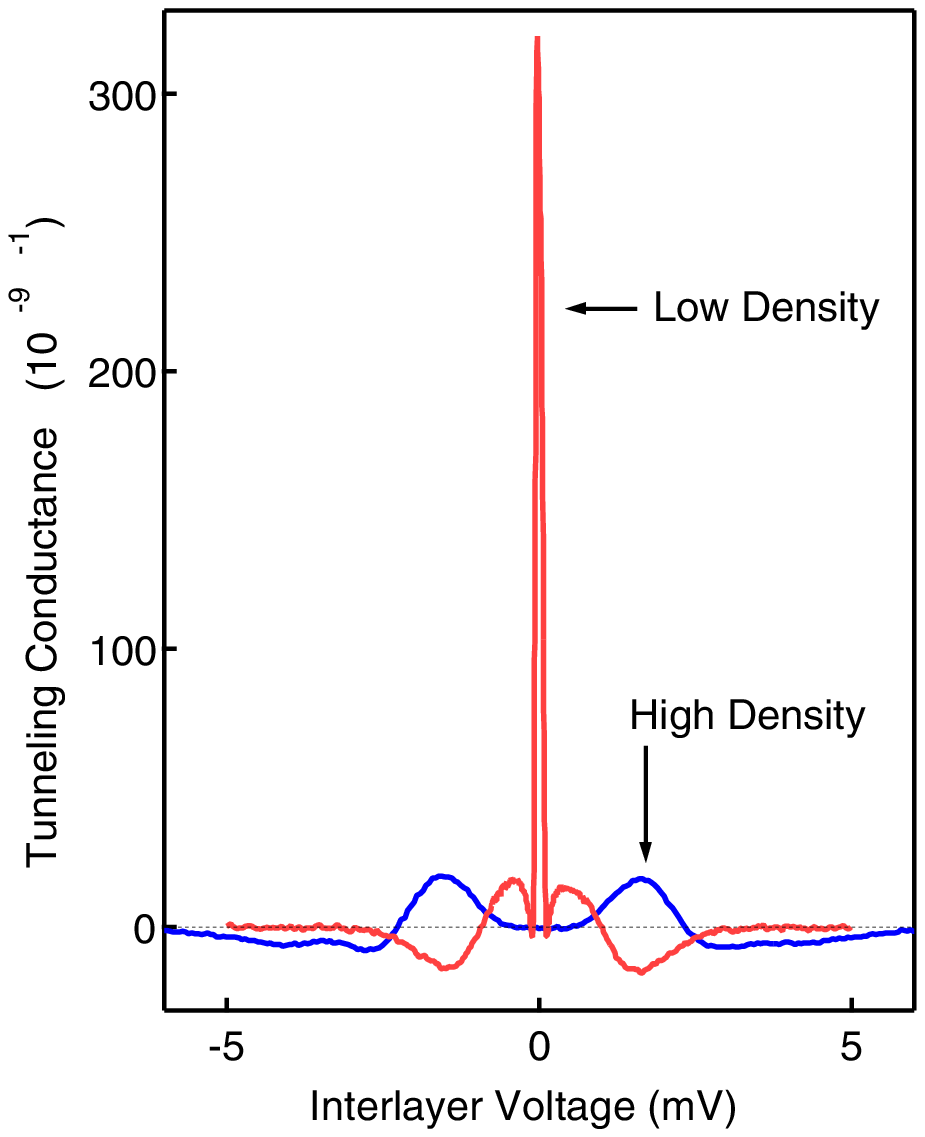} }
 \caption{Upper left panel:  Differential conductance of a QHE bilayer
 system in the phase coherent state at filling factor $\nu=1$ and
 layer spacing $d/\ell=1.61$.  The central peak is remarkably
 narrow with a FWHM of only about $6\mu$eV.  Lower left panel:  $IV$
 curve showing the nearly vertical `supercurrent' branch and a
 remnant of the coulomb gap feature at larger voltages. 
After Spielman et al. Ref.~\citelow{jpegoldstone}
Right panel:
 Differential conductance at low density (small $d/\ell$) in
 the phase coherent state and high density (large $d/\ell$) where the layers
 are uncorrelated.  In the latter case the tunnel current vanishes at
 small voltages due to the Coulomb gap. There is a peak in the current at a
 voltage corresponding to the scale of the Coulomb interactions in the
 system. Figure courtesy of J.~P.~Eisenstein. \label{fig:tunnel1} }
\end{figure}

An ideal experimental probe of the novel properties of the
interlayer phase coherent state is quantum tunneling of electrons
through the barrier separating the two layers via the weak
perturbation \be H_{\rm T} = -\frac{\Delta_{\rm SAS}}{2} \int
d^2r\, \Psi^\dagger_\uparrow(\vec  r) \Psi_\downarrow(\vec  r) +
{\rm h.c.}, \ee where $\Delta_{\rm SAS}$ is the
symmetric-antisymmetric single particle tunnel splitting. Because
this perturbation changes the charge difference between the
layers, it does not commute with rest of the Hamiltonian and one
would naively expect that it would produce an energy shift in the
ground state only in second order perturbation theory.  However
we see from Eq.~(\ref{eq:orderparameter}) that through the magic
of spontaneously broken symmetry, the tunneling term actually has
a finite expectation value in the ground state \be \langle H_{\rm
T} \rangle =  -n_0\frac{\Delta_{\rm SAS}}{2} \int d^2r\,
\cos\varphi(\vec r). \label{eq:tunnel} \ee Thus the response to
tunneling appears in first order, because of the
uncertainty over which layer each electron is in.  The finite
expectation value of the order parameter tells us that we can
transfer an electron from one layer to the other and still be in
the exact same quantum state!  If the system is in the same
quantum state, the energy change is zero.  Therefore the process
conserves energy only if the bias voltage across the tunnel
junction is {\em zero}.  Thus, much as in a superconducting
Josephson junction, we expect an enormous zero bias anomaly in
the tunnel current. \cite{wenzee,ady2PRL,leonleoPRL,fogler} This
prediction, first made by Wen and Zee \cite{wenzee} on the basis
of the broken symmetry ground state proposed by Fertig
\cite{fertig} was recently dramatically confirmed in a remarkable
set of experiments by Spielman et al.
\cite{jpetunnel1,jpegoldstone}  The data shown in
Fig.~(\ref{fig:tunnel1})
represent the differential conductance (and in the lower panel
the conductance) as a function of bias voltage in a sample with
extremely weak tunneling ($\Delta_{\rm SAS} \sim 85\mu$K).

For large layer separation relative to electron spacing, the two
layers are uncorrelated and it costs a lot of Coulomb energy to
suddenly inject an electron into a layer by tunneling. The strong
magnetic field destroys the fermi liquid state by making the
kinetic energy degenerate and preventing the other electrons from
moving away from the newly arrived electron.  Only if the voltage
is large enough to overcome the Coulomb gap will current be able
to flow.  This can be seen in the right panel of Fig.~(\ref{fig:tunnel1}).
Note that if the voltage is {\em too} large the current again
decreases because the excess energy from the bias supply can not
be absorbed by converting it to kinetic energy of the electrons.

For small $d/\ell$ the system undergoes a quantum phase transition
into the interlayer phase coherent state and tunneling is
dramatically enhanced near zero voltage.  Unlike the true
Josephson effect the dissipation is not infinitesimal on the
supercurrent branch.  Various proposals involving a finite phase
coherence time have been made to explain the finite height and
width of the differential conductance peak.
\cite{ady2PRL,leonleoPRL,fogler,yogesh} but this is a question
which is still poorly understood and is a subject of current
study.  In order to understand the finite dissipation, it is
necessary to understand the excitations above the phase coherent
ground state.  We now turn to this question.

Because of the U(1) symmetry, the energy of the coherent state
can not depend on the global phase angle $\varphi$.  Physically
this just results from the fact that, in the absence of tunneling
it is impossible for the electrons to know what the global phase
angle is.  However symmetry does not prevent the energy from
depending on spatial gradients of $\varphi$.   When the
fluctuations out of the easy-plane are small we can parametrize
the local pseudospin orientation vector
$(\cos\varphi,\sin\varphi,m_z)$ in terms of the phase angle
$\varphi$ and the conjugate `charge' $m_z$. The leading terms in
the gradient expansion for the energy yield \be H = \int d^2r\,
\frac{1}{2}\rho_{\rm s} |\nabla\varphi|^2 + \frac{(\hbar
n_0/2)^2}{2\Gamma} m_z^2, \ee where $\Gamma$ is related to the
capacitance per unit area. The pseudospin stiffness $\rho_{\rm
s}$ physically arises from the loss of optimal Coulomb exchange
in the presence of a phase gradient. \cite{leshouches}  Because
momentum in a magnetic field is related to position in real space
\cite{leshouches} a phase gradient of the order parameter
corresponds to a spatial displacement of the correlation hole so
that it is no longer directly on top of the particle.   This
increases the Coulomb energy. Microscopic Hartree-Fock
calculations find $\rho_{\rm s}\sim 0.2-0.5$K for typical sample
parameters.

Because the phase is conjugate to the conserved `charge' $m_z$,
the action takes the form \be S = \int d^2r \int dt\,
[\frac{\hbar n_0}{2} \dot \varphi m_z - H]. \ee The first term
can either be viewed as the Berry phase appropriate to a spin
model \cite{leshouches} or simply as the statement that the
momentum density conjugate to $\varphi$ is $p_\varphi =
\frac{\hbar n_0}{2} m_z$. Integrating out the massive $m_z$
fluctuations yields \be S = \frac{1}{2} \int d^2r \int dt\,
\left[\Gamma \dot\varphi^2 - \rho_{\rm s} |\nabla\varphi|^2
\right]. \ee This is the action of a superfluid with Goldstone
mode velocity $u = \sqrt{\frac{\rho_{\rm s}}{\Gamma}}$.

Because the `charge' $m_z$ conjugate to $\varphi$ is the
difference in the two layer charge densities, the supercurrent
$ \vec J_- = \rho_s \vec
\nabla \varphi $ 
associated with the Goldstone mode 
is antisymmetric in the layer index;  that is,
it corresponds to electronic currents flowing in opposite
directions in the two layers. At finite temperatures the
statistical mechanics of this novel superfluid is that of the 2D
XY model and so it undergoes a true phase transition when
vortices in the order parameter field unbind at the
Kosterlitz-Thouless temperature (on the scale of $\rho_s$). While
there does not yet exist direct evidence for this transition, we
note that the tunnel peak at zero voltage begins to turn on at
temperatures which are consistent with Hartree-Fock estimates of
$\rho_s \sim 100-500$mK.

We are now in a position to analyze the excitations produced by
tunneling.  From Eq.~(\ref{eq:tunnel}) we see that the tunneling
operators that correspond to the two possible directions of
tunneling must be $ T_\pm = - \lambda \int d^2r\,
e^{i\varphi(\vec r)} e^{\pm iQ_{\rm B}x}$ where $\lambda =
n_0\frac{\Delta_{\rm SAS}}{4}$ and the last term allows for the
possibility that there is a tilted magnetic field which puts flux
between the two layers.  This can be derived by choosing the
gauge $\vec A_\parallel = x B_\parallel\hat z$ to describe the
in-plane field and setting $Q_B = \frac{edB_\parallel}{\hbar c}$.

In the true Josephson effect the current is first order in the
tunneling amplitude.  We will assume that there is sufficient
decoherence (to be discussed below) that this does not occur
here.  The linear-response current can then be computed
perturbatively in the tunneling using Fermi's Golden Rule.  For a
sample of size $L^2$ the net tunnel current is
\cite{ady2PRL,leonleoPRL,fogler} \be I(V) = \frac{2\pi e\lambda^2
L^2}{\hbar}[S(Q_B,eV) - S(-Q_B,-eV)], \ee where,
$S(q,\hbar\omega)$ is the spectral density for the fluctuations of
the order parameter at wavevector $q$ and frequency $\omega$, is
proportional to the Fourier transform of the pseudospin
correlation function $\langle e^{i\varphi(\vec
r,t)}e^{-i\varphi(0,0)}\rangle$. Decoherence effects are included
phenomenlogically by adding spatial and temporal damping factors
to this Fourier transform.  \cite{ady2PRL,leonleoPRL}

The current will have exhibit a peak (or in the case of the
differential conductance, a derivative feature) which will occur
at the voltage corresponding to the Goldstone mode energy and
will disperse outwards with increasing wavevector (B field
tilt).  As Fogler and Wilczek \cite{fogler} have noted, this
experiment is closely analogous to the corresponding experiment in
Josephson junctions of Eck et al. \cite{eck} There a feature
appears in the IV characteristic when the wave vector and energy
matching conditions for the Swihart mode are achieved.

\begin{figure}[ht]
\centerline{
\epsfxsize=2.5in \epsffile{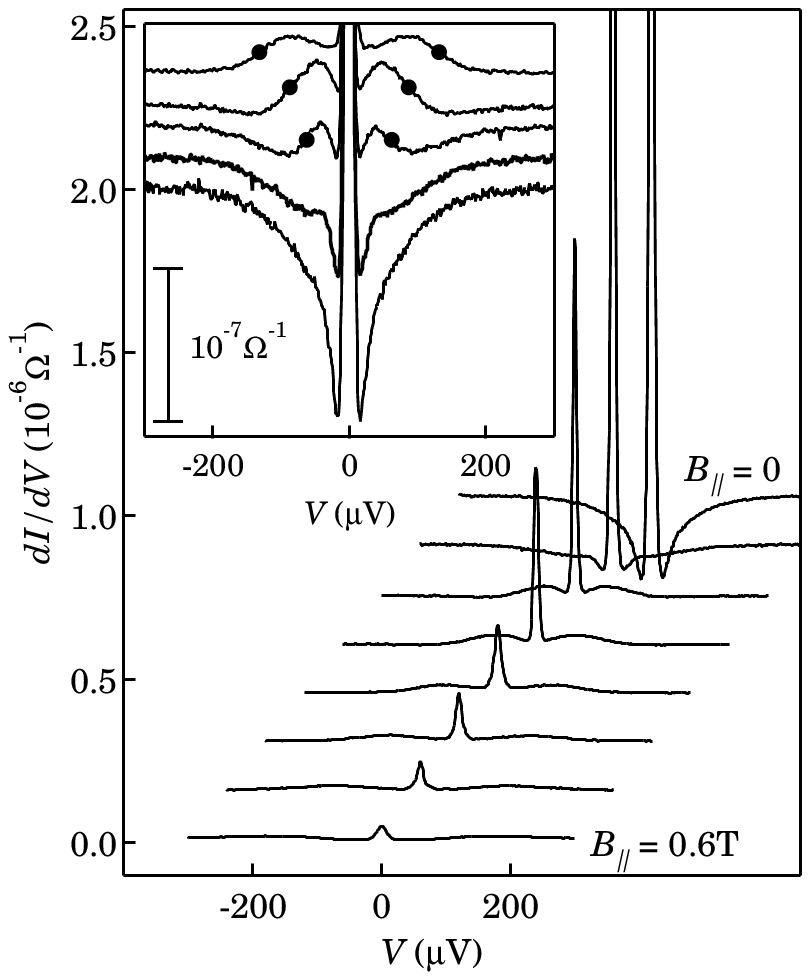}
\epsfxsize=2.5in \epsffile{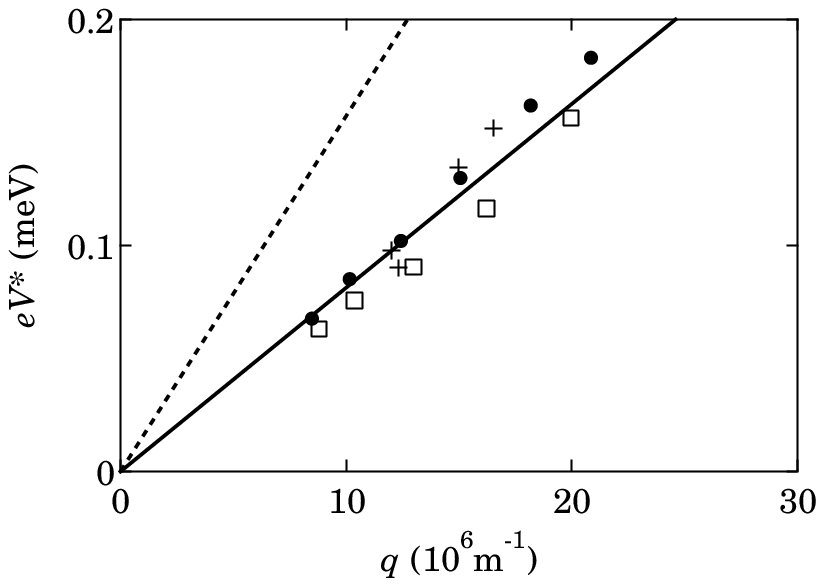}
}
\caption{
Left panel:
Differential conductance vs.~voltage for a variety of
values of the parallel B magnetic $B_\parallel$.  Inset:
Magnified view showing the Goldstone feature dispersing outward
in voltage with increasing $B_\parallel$.  Black dots indicate
inflection points which are used to determine the mode dispersion
shown in the right panel.  The velocity agrees to within about
a factor of two of the value $\sqrt{\frac{\rho_s}{\Gamma}} \sim
10^4$m/s predicted from Hartree-Fock estimates of the spin
stiffness and the compressibility parameters $\rho_s$ and
$\Gamma$. After Spielman et al. Ref.~\citelow{jpegoldstone} }
\label{fig:tunnel2}
\end{figure}

The differential conductance as a function of voltage at
different values of $B_\parallel$ obtained by Spielman et
al.~\cite{jpegoldstone} is shown in the left panel of Fig.~(\ref{fig:tunnel2}).
In two dimenions at finite temperature there is no true broken
symmetry, but order parameter fluctuations still have long-range
power law correlations.  This is the origin of the enormous peak
in the  differential conductance observed for $Q_B=0$. At the
lowest temperatures the half-width of this peak is only
$\delta\sim3\mu$V. Within the perturbative model for the
tunneling line shape this implies a phase coherence time
$\tau_\varphi =\hbar/\delta \sim 2\times 10^{-10}$s.  Using the
collective mode velocity to obtain a measure of the coherence
length yields $\xi \sim 2\mu$m.  This is on the same scale as
both the quantum to classical crossover length $\xi_T \sim \hbar
u/k_{\rm B} T$ and the Josephson penetration length
\cite{ady2PRL,leonleoPRL} $\lambda_{\rm J} =
\sqrt{4\pi\ell^2\rho_s/\Delta_{\rm SAS}}$.  The latter might hint
that the assumption that we can work perturbatively in the
tunneling amplitude may be beginning to break down at the lowest
temperatures.  For finite voltage and finite tunneling the
non-perturbative treatment is quite difficult because the system
is not in equilibrium.  Fogler and Wilczek have attacked this
question in a 1D model by solving the classical equations of
motion for the order parameter. \cite{fogler}

The left panel of Fig.~(\ref{fig:tunnel2}) shows that application of the parallel
magnetic field fairly quickly kills the central peak and a small
side feature appears which disperses outward with increasing
$B_\parallel$.  Spielman et al.~identify the inflection point as
the center of this derivative feature and plot the resulting
dispersion curve as shown in the right panel of Fig.~(\ref{fig:tunnel2}).
The dispersion is indeed linear and agrees to within about a
factor of two of the predicted mode velocity of $\sim 10^4$m/s.
It is perhaps not surprising that the measured mode velocity is
somewhat lower since quantum fluctuations neglected in the
Hartree-Fock approximation will lower the spin stiffness.

In addition to tunneling, another good probe of the phase
coherent state would be interlayer drag.
\cite{kmoonprb,kyangdrag,kimdrag} In a drag experiment one uses a
sample with negligible tunneling, drives a current through the
upper (say) layer and then measures the voltage drop in the lower
layer (under zero current conditions for that layer).  The ratio
of drag induced voltage to the applied current is called the
transresistance.  For the case of ordinary fermi liquids, the
electric field in the lower layer is directed {\em oppositely} to
that in the drive layer in order to counteract the drag force due
to momentum transferred from the drive layer.  Because the rate
of collisions between the fermi liquid particles vanishes as
$T^2$ at low temperatures, the drag resistance is small and
vanishes at zero temperature.

The simplest way to analyze drag in the present case is to define
symmetric and antisymmetric (in layer index) currents $ \vec
J_\pm = \vec J_\uparrow - \vec J_\downarrow$.  Transport in the
symmetric channel is that of an ordinary, nearly dissipationless,
$\nu=1$ Hall plateau.  However transport in the antisymmetric
channel is that of a superfluid.  The antisymmetric electric
field must therefore {\em vanish}.  This means that the electric
field in the lower layer must exactly match that in the drive
layer.  Hence the drag is very large (and does not vanish at
$T=0$)  and the transresistance tensor $\rho^{12}$ is equal to
the quantized Hall resistance tensor
\cite{kmoonprb,kyangdrag,chetan}
$\rho_{xx}^{12} \approx 0;\,\,\,\,\,\,\,
\rho_{xy}^{12} = \frac{h}{e^2}$.

At the present time, there are still a variety of open issues.
We do not have a microscopic understanding of the
dissipation/decoherence mechanism which gives a finite width and
height to the tunneling peak. \cite{yogesh}
  We do not
understand why the central peak is not destroyed more rapidly
with the addition of $B_\parallel$.  The peak is still visible
even when the Goldstone feature has moved out far enough to be
distinct from it.  Finally, we do not have a good understanding
of the nature of the quantum phase transition or transitions that
occur as the layer spacing is increased. Various scenarios have
been suggested theoretically \cite{bonesteel,chetan,veillette}
and there is some numerical evidence hinting that there is a single
weakly first order transition.  \cite{Schlieman}

\section*{Acknowledgments}
This work was supported by NSF DMR-0087133 and represents
long-standing collaborations with many colleagues including Allan
MacDonald, Ady Stern, J. Schlieman, Ning Ma, K. Moon, and K.
Yang.  I also would like to thank J. P. Eisenstein and his group
for numerous helpful discussions of their experiments.

\end{document}